\documentclass[twocolumn,amsmath,amssymb,showpacs,superscriptaddress,prl,aps,longbibliography,10pt]{revtex4-1}
\usepackage{graphicx}
\usepackage{bm}
\usepackage{dcolumn}
\usepackage{amssymb}
\usepackage{amsmath}
\usepackage{amsfonts}
\usepackage{epsfig}
\usepackage{graphicx}
\usepackage{stackrel}
\usepackage{paralist}
\usepackage[rgb]{xcolor}
\usepackage{todonotes}
\usepackage{pdfcomment}
\usepackage{mathrsfs}

\newcommand{\I}{\mathrm{i}}
\newcommand{\refEq}[1]{Eq.~(\ref{#1})}
\newcommand{\refFig}[1]{Fig.~\ref{#1}}

\newcommand{\abs}[1]{\left\vert#1\right\vert}

\begin{document}
\title{Setting Boundaries with Memory: Generation of Topological Boundary States in Floquet-Induced Synthetic Crystals
}
\author{Yuval Baum}
\affiliation{Institute of Quantum Information and Matter, California Institute of Technology, Pasadena, California 91125, USA}
\author{Gil Refael}
\affiliation{Institute of Quantum Information and Matter, California Institute of Technology, Pasadena, California 91125, USA}
\begin{abstract}
When a $d$-dimensional quantum system is subjected to a periodic drive, it may be treated as a $(d+1)$-dimensional system, where the extra dimension is a synthetic one. In this work, we take these ideas to the next level by showing that non-uniform potentials, and particularly edges, in the synthetic dimension are created whenever the dynamics of system has a memory component. We demonstrate that topological states appear on the edges of these synthetic dimensions and can be used as a basis for a wave packet construction. Such systems may act as an optical isolator which allows transmission of light in a directional way. We supplement our ideas by an example of a physical system that shows this type of physics.  
\end{abstract}
\maketitle

\emph{Introduction.}
The discovery of new phases of matter and the ability to the manipulate them are the essence of condensed matter physics. While solid-state or cold-atom systems offer many paths to realizing novel quantum phases,       
in recent years, several new means of realizing interesting quantum phases have been proposed. Two promising proposals in this direction are synthetic dimensions and periodic drives. Synthetic dimensions constitute a reinterpretation of discrete internal degrees of freedom that play the role of lattice sites, and hence of additional dimensions. 
Several physical realizations of synthetic dimensions were put forward  \cite{SD1,SD2,SD3,SD4,SD5,SD6}. Among them are ultra-cold gases \cite{SD_CA1,SD_CA2,SD_CA3}, where the synthetic dimension is implemented by employing internal atomic states, and optical systems \cite{SD_reso1,SD_reso2,SD_reso3,SD_reso4}, in which the modes of a ring resonator at different frequencies take the role of the lattice sites.    
Periodic drives have been also proposed as a tool for generating new phases. They may alter the electronic spectrum of crystals \cite{Fl_exp1,Fl_exp2,Fl_exp3,Fl_g1,Fl_g2,Fl_g3,Fl_g4,Fl_g5,Fl_g6,Fl1,Fl2,Fl3,Fl4,Fl5,Fl6,mbl1,mbl2,mbl3,TC1,TC2,TC3,TC4}, leading to exotic phases and phase transitions, among them, the topological and Anderson Floquet insulators \cite{Fl1,Fl2,Fl3,Fl4,Fl5,Fl6}, time crystals \cite{TC1,TC2,TC3,TC4} and the many body localization-delocalization transition \cite{mbl1,mbl2,mbl3}.

Periodic drives may also modify the system by introducing a synthetic dimension. Within the Floquet framework, quantum states become dressed by all possible harmonics of the drive frequency (photons). As used in Ref. \cite{two_drives}, the number of photons (i.e., the harmonic) appearing in a Floquet dressed state may serve as a synthetic dimension. Such a link could be a powerful control tool, since it allows introducing additional dimensions which are externally controlled, and it does not rely on the internal structure of the quantum system considered. 

These, so-called, Floquet synthetic dimensions, however, come at a price.  The ubiquitous time derivative in the Shcr\"{o}dinger equation results in a linear potential, and hence a force, along the Floquet dimension. This field tends to conceal the desired effect of the increased dimensionality, and could even be the dominant energy scale in the problem. In addition, the Floquet synthetic dimension is always translationally-invariant, i.e., only hopping and uniform on-site terms are allowed \cite{trans_com}. Given that topological behavior is most conspicuously exhibited on edges, their absence is particularly limiting when exploring topological phases in synthetic dimensions.  

In this work, we seek to overcome these limitations by going beyond the standard Floquet framework.
The key insight is that if a system's dynamics depends on its past, it could bypass the two restrictions of Floquet synthetic dimensions. Just as real-space potentials correspond to mixing different momentum states, non-uniform potentials in the Floquet space correspond to non-diagonal elements in the time domain. 
Below, we demonstrate how non-locality in-time, and memory effects in particular, allow the control of the effective potential along the Floquet synthetic dimensions. Not only could the undesirable electric fields be eliminated, but edges in the synthetic dimension can also be created. 
We apply this idea to zero and one-dimensional synthetic-dimension topological systems, explain how such memory dependence could be constructed, and discuss possible applications.  
\begin{figure}[t]
\centering
\includegraphics[width = \linewidth]{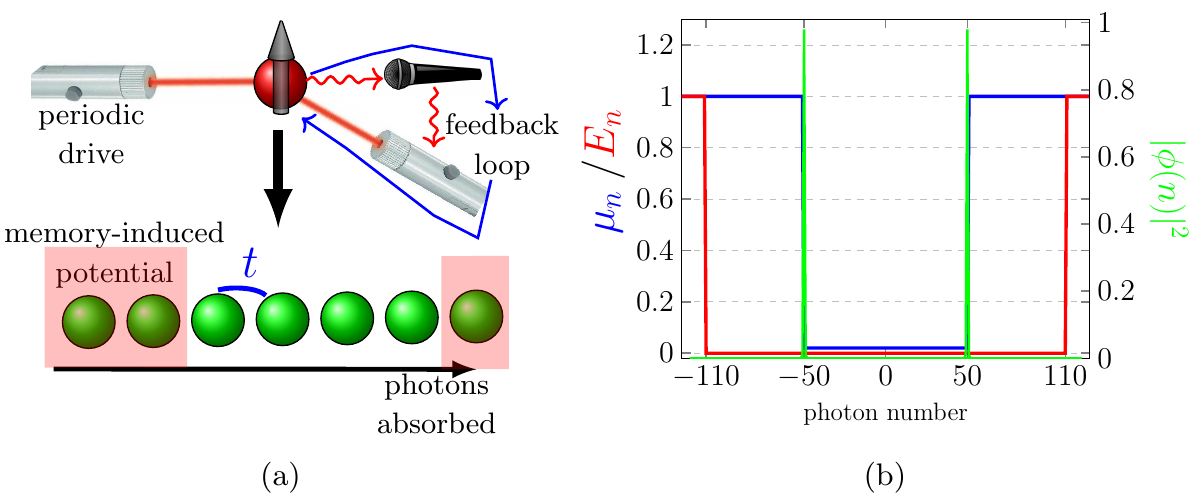} 
\caption{(a) Cartoon of the equivalence between a two level system in the presence of an external drive and a memory kernel and particles hopping on a $1$D synthetic dimension lattice. (b) The $n$-dependent gap parameter $\mu$ as defined in \refEq{Eq:H} (blue), the $n$-space effective electric field in units of $5\hbar\omega$ (red) and The absolute square of the two solutions of \refEq{Eq:Sch3} with $\eta=0$ (green). Evidently, there is an electric-field-free zone in $n$-space in which the system hosts two localized solutions near the jump in the gap parameter. Here, $n_0=110$, $n_1=50$, $\mu=10$ and $\delta=-9.9$. \label{WF}}
\end{figure}
\emph{0+1 dimensional model.}
Our first goal is to map a the dynamics of a periodically-driven quantum system, including memory effects, into a lattice model. 
Consider the non-Markovian evolution of such a quantum system: 
\begin{equation}\label{Eq:Sch1}
\I\partial_t\psi(t)=\mathcal{H}(t)\psi(t)+\int_{0}^{\infty}{\mathcal{U}(\tau)\psi(t-\tau)\mathrm{d}\tau},
\end{equation}
where $\mathcal{H}(t)=\mathcal{H}(t+T)$ is a time periodic Hamiltonian and $\mathcal{U}$ is a memory kernel that captures the non-Markovian effects in the system. 
Since \refEq{Eq:Sch1} is invariant to time translations by $T$, its solutions have a Floquet form,
\begin{equation}\label{Eq:floq_form}
\psi(t)=e^{-\I\eta t}\sum\limits_{n=-\infty}^{\infty}\phi_ne^{\I n\omega t},
\end{equation}
where $\omega=2\pi/T$, and $\phi_n$ is the Floquet amplitude for an electronic state dressed by $n$ photons. In contrast to the Markovian case, \refEq{Eq:Sch1} does not preserve the norm of $\psi$, and therefore $\eta$ is a general complex number. 

\refEq{Eq:floq_form} and \refEq{Eq:Sch1} yields the following equation for the Floquet amplitudes,
\begin{equation}\label{Eq:Sch2}
\eta\phi_n=\big(\omega n+F_n(\eta)\big)\phi_n+\sum\limits_{m}\mathcal{H}_{n-m}\phi_m,
\end{equation}
where $F_n(\eta)=\int_{0}^{\infty}{\mathrm{d}\tau \mathcal{U}(\tau)e^{\I\tau(\eta-n\omega)}}$ and $\mathcal{H}_{n}=\int_0^T{\frac{\mathrm{d}\tau}{T}\mathcal{H}(t)e^{-\I n\tau\omega}}$.
Unlike standard tight-binding models, \refEq{Eq:Sch2} is a transcendental eigenvalues equation and the number of independent solutions depends on the exact form of $F_n$ and $\mathcal{H}$.

For simplicity and clarity, we restrict our discussion to memory kernels of the form: $\mathcal{U}(t)=\Theta(t)\Theta(T-t)\,u(t)/T$, where $\Theta$ is the Heaviside function, and $u(t)$ is periodic in t with period $T$. This form implies that the memory is causal and goes back only up to a single period of the drive. In this case the memory kernel is fully defined by $u_n$, the Fourier components of $u(t)$ in the range $t\in[0,T]$.
In terms of $u_n$, $F$ is given by:
\begin{equation}\label{Eq:F}
F_n(\eta)=\sum_l\frac{u_l}{T}\int_0^T{e^{\I\tau(\eta+(l-n)\omega)}\mathrm{d}\tau}.
\end{equation}
Notice that $F_n(\eta=0)=u_n$. $u_n$ plays the role of a potential energy in photon space. The scale of $u_n$ characterizes the coupling of the system to its memory. 

Consider a concrete problem of a two-level system in the presences of a single frequency drive. From this system, we will construct a $1$D topological phase. Start with: 
\begin{equation}\label{Eq:H}
\mathcal{H}(t)=\big(\mu-\cos{(\omega t)}\big)\sigma_z+\sin{(\omega t)}\sigma_x,
\end{equation}
here $\sigma_i$ are the Pauli matrices and $\mu$ is a positive parameter.
We choose the memory kernel such that \cite{kernel}:
\begin{equation}\label{Eq:un}
u_n=-\omega n\Theta(n_0-|n|)\sigma_0+\delta\Theta(n_1-|n|)\sigma_z,
\end{equation}
where $\sigma_0$ is the identity matrix and $\delta$ is a real parameter.
In order to compete with the artificial electric field, we see that the memory coupling  should be of order $\omega N$, where $N$ is the extent of the flat region we would like to have. In addition, $\delta$ sets the scale of the $n$-space confinement potential, which in turn, sets the extent of the boundary states. The smaller this extent is, the better defined wave-packets could be created along the edge.
From \refEq{Eq:H}, \refEq{Eq:un} and \refEq{Eq:Sch2} we find,
\begin{align}\label{Eq:Sch3}
\eta\phi_n=&\Big(\omega n\Theta(|n|-n_0)\sigma_0+\tilde{F}_n(\eta)\Big)\phi_n\\ \nonumber
&+\mu_n\sigma_z\phi_n-\frac{\sigma_z+\I\sigma_x}{2}\phi_{n+1}-\frac{\sigma_z-\I\sigma_x}{2}\phi_{n-1},
\end{align}
where $\tilde{F}_n(\eta)=F_n(\eta)-u_n$ and $\mu_n$ is $\mu+\delta$ for $|n|\leq n_1$ and $\mu$ otherwise.

The last three terms in \refEq{Eq:Sch3} describe a $1$D Kitaev chain \cite{Kitaev} with a jump in the topological mass at $n=\pm n_1$.
For $|\mu|>1$ and $|\delta+\mu|<1$, the region between $n=\pm n_1$ is in the topological phase while the exterior is in a trivial phase. Overall, the spectrum arising from these three terms alone is gapped along with two zero energy states which are localized in $n$ space around $n=\pm n_1$.
The first term in \refEq{Eq:Sch3} describes a constant electric field which is perfectly screened in the region $|n|<n_0$. For $n_1\ll n_0$, the low energy states are indifferent to that field, and we may assume for simplicity that $n_0\to\infty$, i.e., the electric field is perfectly screened.
In that limit and for $\eta=0$ and $\tilde{F}_n(\eta)=0$, \refEq{Eq:Sch3} describes a zero energy Kitaev chain which, as explained above, has two solutions which are localized in $n$ space around $n=\pm n_1$.
Indeed we find two solutions only which are localized at the edges of the synthetic direction. The $n$ space wave functions of these two solutions are depicted in \refFig{WF}b, and they are given by:
$\psi_{\pm}(t)=\sum\limits_{n=-\infty}^{\infty}\phi_n e^{\I n\omega t}\approx \phi_{\pm n_1}e^{\pm\I n_1\omega t}.$
For a general $\eta=x+\I y\neq 0$, the spectrum of the right-hand-side of \refEq{Eq:Sch3}, $\lambda_i$, may be found for each value of $\eta$. Only eigenstates with $\eta=\lambda_i$ are true solutions of \refEq{Eq:Sch3}.
We verified numerically that near zero the two solutions above with $\eta=0$ are the only solutions. 

This example demonstrates how to construct a synthetic dimension using the photon number as a lattice degree of freedom. The inclusion of a memory kernel allowed us to introduce potentials, and in particular, edges in the synthetic dimension. Unlike standard lattice models, here, only the zero energy states are eigenfunctions of the dynamics. Next, we add another dimension to find a collection of eigenfunctions of the dynamical equation.

\emph{1+1 dimensional model.} Next, let us construct a $2$D model which consists of one real and one synthetic dimension. We start with:
\begin{equation}\label{Eq:Sch2D1}
\I\partial_t\psi_x(t)=\sum\limits_{x'}\mathcal{H}_{x-x'}(t)\psi_{x'}(t)+\int\limits_0^T{\frac{\mathrm{d}\tau}{T}u_{x-x'}(\tau)\psi_{x'}(t-\tau)},
\end{equation}
where $x$ is a lattice coordinate, $\mathcal{H}(x-x',t)$ is a time-periodic tight-binding Hamiltonian and $u$ is the memory kernel.
For periodic or infinite systems in the real dimension, \refEq{Eq:Sch2D1} can be written in Fourier space:  
\begin{equation}\label{Eq:Sch2D2}
\I\partial_t\psi(k,t)=\mathcal{H}(t,k)\psi(k,t)+\int_0^T{\frac{\mathrm{d}\tau}{T}u(k,\tau)\psi(k,t-\tau)}.
\end{equation}
We choose $u(k,\tau)=u(\tau)e^{-\I v_0k\tau}$, where $u(\tau)$ is again given by \refEq{Eq:un}, and $v_0$ is a real parameter. As before, the solution has a Floquet form and \refEq{Eq:Sch2D2} becomes: 
\begin{equation}\label{Eq:Sch32D}\nonumber
\eta_k\phi_n(k)=\big(\omega n+F_n(\eta_k-v_0k)\big)\phi_n(k)+\sum\limits_{m}\mathcal{H}_{n-m}(k)\phi_m(k)
\end{equation}
with a similar $F_n$ as in the previous section. 

Next, we generalize the Hamiltonian of the zero dimensional case \cite{Chern1,BHZ}: 
\begin{equation}\label{Eq:H2D}\nonumber
\mathcal{H}(t)=\big(\mu-\cos{(\omega t)}-\cos{(k)}\big)\sigma_z+\sin{(\omega t)}\sigma_x+v_0\sin{(k)}\sigma_y.
\end{equation}
Taking the limit $n_0\to\infty$ and trying solutions with $\eta_k=\pm \abs{v_0}k$ yields the following eigenvalues equation,
\begin{align}\label{Eq:Sch52d}\nonumber
\pm v_0k\phi_{n,k}&=\Big[\big(\mu_n-\cos(k)\big)\sigma_z+v_0\sin{(k)}\sigma_y\Big]\phi_{n,k}\\ 
&-\frac{\sigma_z+\I\sigma_x}{2}\phi_{n+1,k}-\frac{\sigma_z-\I\sigma_x}{2}\phi_{n-1,k}.
\end{align}
The right hand side of \refEq{Eq:Sch52d} is a tight-binding model of a Chern insulator on a cylinder in the $x-n$ plane. The Chern number changes from $1$ to $0$ at $n=\pm n_1$. Similar to a quantum Hall (QH) state on a cylinder, there are no low-energy states in the $2$D bulk, while for each $k$, two chiral solutions with energy $\epsilon_k=\pm v_0k$ exist near the edges at $n=\pm n_1$.
Hence, we found a set of solutions labeled by $k$:
\begin{equation}\label{Eq:2D_sol}
\psi_{k,\pm}(x,t)=e^{\I kx}\psi(k,t)\approx e^{\I kx\pm\I v_0k t}\phi_{k,\pm n_1}e^{\pm\I n_1\omega t}.
\end{equation}
The solutions may be superposed to construct a wave packet:
\begin{align}\nonumber
\Psi_{\pm}(x,t)=&\sum\nolimits_k\psi_{k,\pm}(x,t)A(k)\propto A(x\pm v_0 t),
\end{align}
where $A(k)$ is an envelope function tightly centered around $k=0$.
Overall, we generated a wave packet with a fixed number of photons that propagates, without dispersion, along the $1$D chain.

\begin{figure}[t!]
\centering
\includegraphics[width =\linewidth]{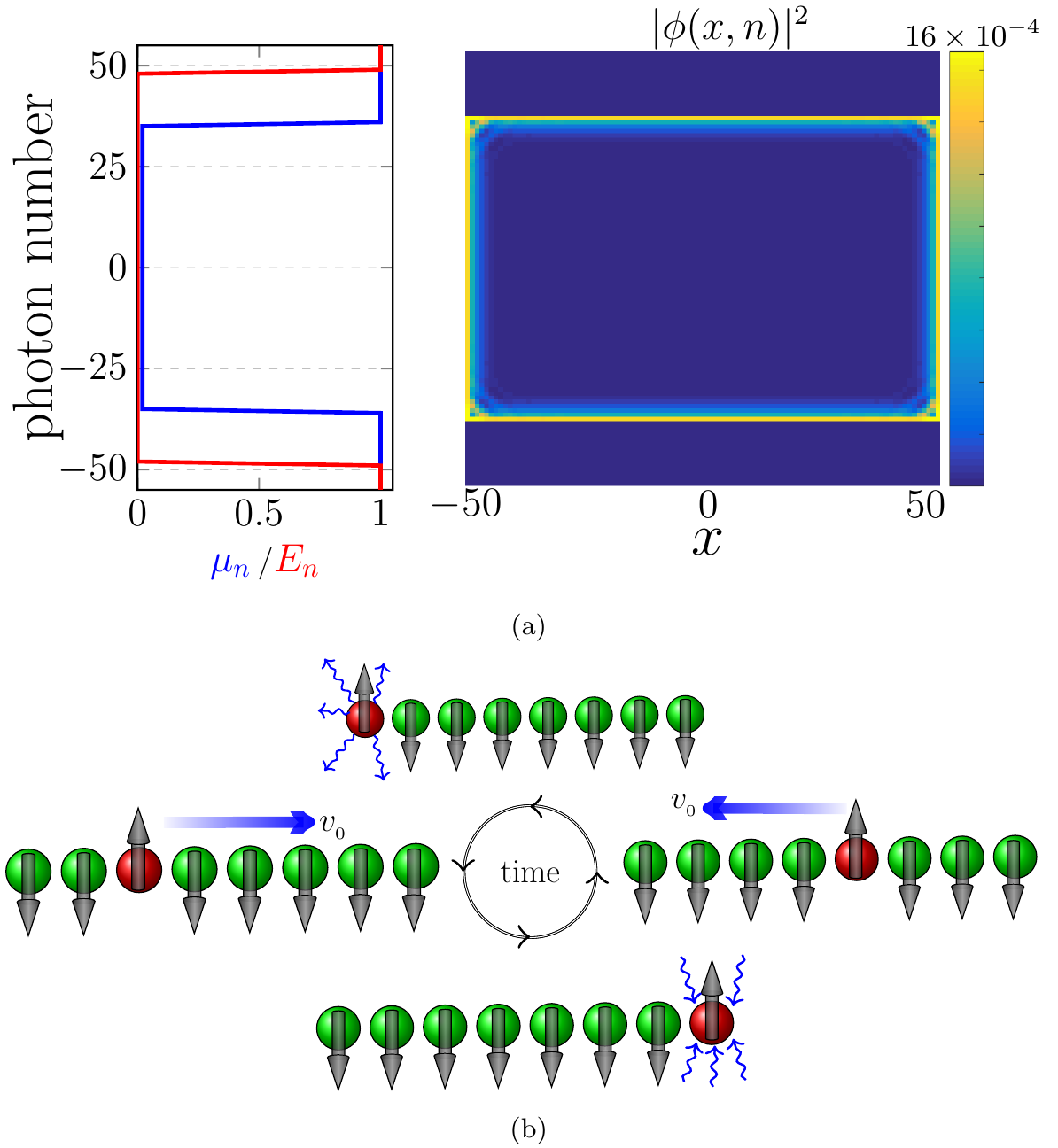} 
\caption{(a) Similar to figure 1, only for the $1+1$D case. The absolute square of a solution to \refEq{Eq:Sch32D} with $\eta=0$ along with $n$ dependent the electric field and gap parameter. At low energies, the system supports only QH-like edge states in the combined $x-n$ space. Here, $n_0=48$, $n_1=35$, $\mu=10$ and $\delta=-9.9$. (b) Cartoon of the cycle that a system of spinful particles performs. Here, spin up (down) denotes full (empty) site. The system evolves along the state which is depicted in (a). Along the left/right real space edge the system emits/absorbs photons while along the top/bottom $n$ space edge the system supports left/right propagation. \label{cycle_demo}}
\end{figure}

We can now add edges in the real direction. \refEq{Eq:Sch32D} can be transformed back to real space and may be solved numerically. As in the periodic case, a set of solutions with real $\eta$ exists. Analogous to a QH state in a rectangular geometry, there are no low-energy states in the $2$D bulk, while a set of chiral solutions, labeled by their energy $\eta$, exists along the circumference of the $2$D sample. Here, the circumference has both real and synthetic segments near the real edges of the chain and $n=\pm n_1$. A typical solution is shown in \refFig{cycle_demo}b, and it is clearly concentrated along the edges of the combined $2$D system.
As expected from the QH analogy, the solutions are propagating plane waves along the circumference. Indeed, we find numerically that the solutions on the left/right real space edges have the form $\phi^{\eta}_{l/r}(n)\propto e^{\pm\I\xi\eta n}$, and therefore, $\psi^{\eta}_{l/r}(t,n)\propto e^{\I\eta(t\pm\xi n)}$ where $\xi$ is a constant that depends on the details of the edge potential.
Constructing a wave packet of different energy solutions around the left/right edge yields,
$
\Psi_{l/r}(n,t)\propto A(\pm \xi n-t),
$
where $A$ is an envelope function tightly centered around zero.
As long as $-n_1<n<n_1$, a wave packet which is localized near the left (right) edge adsorbs (emits) photons from (to) the drive at a constant rate $\xi^{-1}$.

Overall, the wave packet performs a cycle in the $x-n$ plane. If a wave packet with a well defined number of photons, $n_1$, is prepared in the bulk of a $1$D chain, then it moves at a constant velocity and without dispersion toward the right edge. At the edge it emits photons at a constant rate until it reaches $n=-n_1$. It then moves toward the opposite edge where it absorbs back photons from the drive until $n=n_1$ again, and therefore completes a cycle. \refFig{cycle_demo}c illustrates such a cycle for a chain with an internal pseudospin degree of freedom. 
In appendix A we describe a local-in-time framework in which the dynamics of these systems may be simulated efficiently. 

\begin{figure}[t]
\centering
\includegraphics[width =\linewidth]{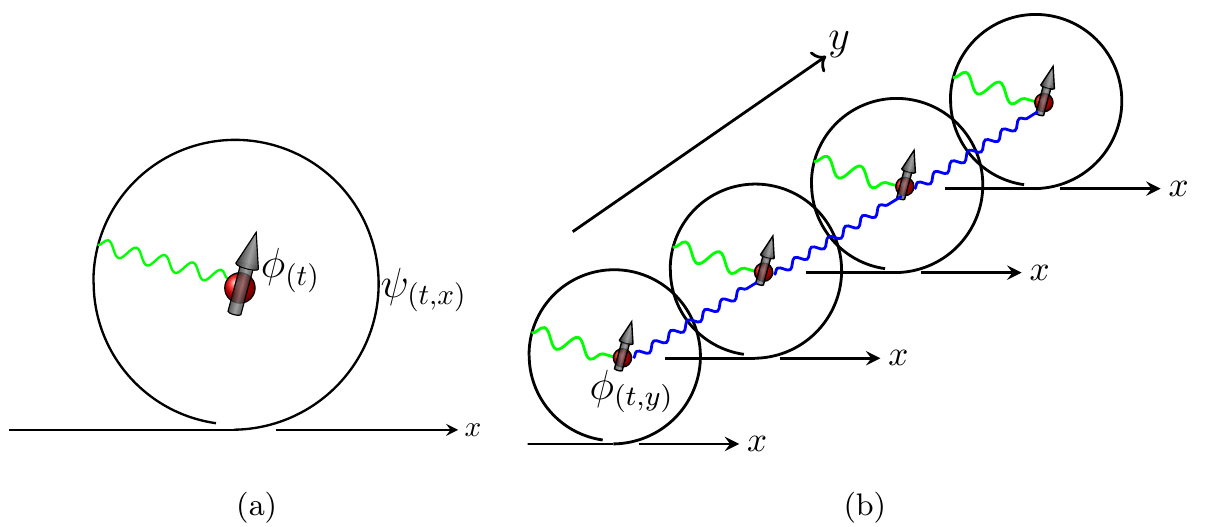} 
\caption{(a) Cartoon of the system in \refEq{Eq:EOM1} and \refEq{Eq:EOM2}. $\phi$ represents a pseudo-spin degree of freedom which is coupled to a chiral field $\psi$. The effective evolution of the $\phi$ field follows \refEq{Eq:Sch1}. (b) The one dimensional version of (a). Placing a $1$D lattice of the system in (a) and introducing couplings between the different two-level systems yields an effective $1$D model for the $\phi$ field that obeys \refEq{Eq:Sch2D1}. Each site has a pseudospin degree of freedom that is either full or empty.\label{swirl}}
\end{figure}

\emph{Implementation.} A dynamical equation of the form of \refEq{Eq:Sch1} could be engineered by using an auxiliary subsystem.
Consider a two-level system, described by $\phi(t)$ and governed by $H_{\phi}(t)$, which is coupled to a one-dimensional field, $\psi(x,t)$, which is governed by $H_{\psi}(\hat{x})$ and lives on an infinite line that swirls around the two-level system (\refFig{swirl}a). The fields' dynamics are given by \cite{action}:
\begin{align}\label{Eq:EOM1}
&\Big(\I\partial_t-H_{\phi}(t)\Big)\phi(t)=-\int\mathrm{d}x\,\lambda(x)\psi(x,t)\\ \label{Eq:EOM2}
&\Big(\I\partial_t-H_{\psi}(\hat{x})\Big)\psi(x,t)=-\lambda^{\dagger}(x)\phi(t) 
\end{align}
where the coupling, $\lambda(x)$, is non-zero only for $0<x<2\pi L$ with $L$ being the circumference of the swirl and $x=0$ is the starting point of the swirl.
The second equation may be solved formally by introducing $G_{\psi}$, the Green function of the operator $\I\partial_t-H_{\psi}(\hat{x})$. Plugging the formal solution back yields an equation for $\phi$ similar to \refEq{Eq:Sch1}, where the memory kernel is given by: 
\begin{align}\label{kernel_general_form}
\mathcal{U}(t-t')=\int\mathrm{d}x\mathrm{d}x'\,\lambda(x)\,G_{\psi}(t-t',x-x')\lambda^{\dagger}(x').
\end{align}
For $H_{\psi}(\hat{x})=-\I v_0\partial_x$, the Green function is $G_{\psi}(t,x)=\I\Theta(t)\delta(x-v_0t)$, and therefore,
\begin{align}\label{kernel_real}
\mathcal{U}(t)
\sim\Theta(t)\Theta(T-t)\sum\limits_{n,m\neq n}\frac{2\mbox{Re}(\lambda_{m}\,\lambda_{n}^{\dagger})}{2\pi (n-m)}e^{\I n \omega_0 t},
\end{align}
where $\lambda_l$ are the Fourier components of $\lambda(x)$ in the range $x\in[0,2\pi L]$ and the last equality is up to an additive constant. 
\refEq{kernel_real} has the general form of the memory kernel that we considered in the previous sections. Hence, $\omega_0$ and $\lambda(x)$ may be controlled to produce the desired $u(t)$. In appendix B we discuss specific choices of $\lambda_n$ that yield the desired terms. We also comment on the applicability of the above method for a non-chiral $\psi$ field.
The setup in \refFig{swirl}{a} reproduces the physics of \refEq{Eq:Sch1}. The one dimensional physics of \refEq{Eq:Sch2D1} may be approached by placing a $1$D array of these building blocks and introducing couplings between the different two-level systems as illustrated in \refFig{swirl}(b). In this example, each site has two orbitals corresponding to the spin or pseudospin quantum number, which could be either full or empty.

For a $1$D system, we imagine that at $t=0^-$ the system is prepared such that only the most left two-level system is occupied and the drive is turned on (rapidly) at $t=0$. At time $t=0^+$ the system is in a localized state near the left edge and near $n=0$, and then evolves according to the cycle in \refFig{cycle_demo}.

\emph{Conclusions.} 
We showed that dynamical systems which are governed by the combination of periodic drives and memory kernels, may be mapped into synthetic dimension lattices with broken translation invariance. The drive-induced synthetic dimension can possess edges that could host chiral topological states. Memory assisted dynamics is thus a tool that enhances our control of synthetic Flouqet lattices, which could be used in broader contexts. It could be used to protect quantum states from external noise, or to induce through feedback a desirable target state. 

The use of memory for inducing synthetic dimensions edge states could have applications in controlling the flow of light, and in particular its direction. These are key for integrated optical circuits, as
nonreciprocal optical devices, like optical diodes (isolators) have the potential to largely outperform their electronic counterparts \cite{isolator}. Such devices require time-reversal symmetry breaking. In Faraday , for instance, isolators time-reversal is broken by the existence a magnetic field. Our construction may serve as a frequency-dependent isolator that does not require an external magnetic field. The chiral nature of this phase provides the necessary ingredient, and it emerges from the circular polarization of the drive source. At low energies, the 1D system in \refEq{Eq:Sch2D1} supports states in which photon absorption is possible only along the left edge, photon emission is possible only along the right edge and no emission or absorption are possible in the bulk. By connecting the system to input and output ports and injecting light at frequency which is an integer multiple of $\omega$, the system behaves as an isolator. 
Also, a combination of the surface states of a Weyl semimetal with spin orbit coupled wires could provide a physical realization of such an isolator. We defer a discussion of the specifics of such a system, as well as other potential applications of memory-based quantum control to future work.

\begin{acknowledgments}
\emph{Acknowledgments ---} GR is grateful to the the NSF for funding through the grant  DMR-1040435, the Packard Foundation as well as the Aspen Center for Physics, funded by NSF grant PHY-1607611, where part of the work was done. 
GR and YB are grateful for support through the IQIM, an NSF physics frontier center funded in part by the Moore Foundation. 
\end{acknowledgments}

\appendix
\section{Appendix A: Local-in time Formalism and numerical evolution}
In this appendix we derive a local-in-time formalism for the dynamics of \refEq{Eq:Sch2D1}. This formalism was used to produce the numerical time evolution results shown in the main text. Our starting point is:
\begin{align}\nonumber
\I\partial_t\psi_x(t)=&\sum\limits_{x'}\mathcal{H}_{x-x'}(t)\psi_{x'}(t) 
+\int\limits_0^T{u_{x-x'}(\tau)\psi_{x'}(t-\tau)\frac{\mathrm{d}\tau}{T}}. \nonumber
\end{align}
Since $u(\tau)$ is defined only in the range $\tau\in[0,T]$, we may write it as $u_{x-x'}(\tau)=\sum\limits_n u_{n,x-x'}e^{\I n\omega \tau}$, where $\omega=2\pi/T$. Next, we define the following quantity:
\begin{equation}\label{phi_def}
\phi_{n,x'}(t)=\frac{1}{T}\int\limits_0^T{e^{\I n\omega \tau}\psi_{x'}(t-\tau)\mathrm{d}\tau}.
\end{equation}
Hence, \refEq{Eq:Sch2D1} can be written as follows:
\begin{equation}\nonumber
\I\partial_t\psi_{x}(t)=\sum\limits_{x'}\mathcal{H}_{x-x'}(t)\psi_{x'}(t)+\sum\limits_n u_{n,x-x'}\phi_{n,x'}(t).
\end{equation}
The time derivative of \refEq{phi_def} fulfills:

\begin{align}\nonumber
\I\partial_t\phi_{n,x}(t)&=-\int_0^T{e^{\I n\omega \tau}\partial_{\tau}\psi_{x}(t-\tau)\frac{\I\mathrm{d}\tau}{T}} \\ \nonumber
&=-\frac{\I}{T}\Big[\psi(x,t-T)-\psi(x,t)\Big]-n\omega\phi_{n,x}(t)\\ \nonumber
&=\frac{1}{\I T}\Big[e^{\I\eta T}-1\Big]\psi_{x}(t)-n\omega\phi_{n,x}(t),
\end{align}
where in the last step we assumed that $\psi$ has a Floquet form.
Overall, we replaced an integro-differential equation, \refEq{Eq:Sch2D1}, by a set of local-in-time differential equations:
\begin{align}\nonumber
&\I\partial_t\psi_x(t)=\sum\limits_{x'}\mathcal{H}_{x-x'}(t)\psi_{x'}(t)+\sum\limits_n u_{n,x-x'}\phi_{n,x'}(t) \\ \nonumber
&\I\partial_t\phi_{n,x}(t)=\frac{1}{\I T}\Big[e^{\I\eta T}-1\Big]\psi_x(t)-n\omega\phi_{n,x}(t)
\end{align}
Notice that unlike the set of equations in \refEq{Eq:EOM1} and \refEq{Eq:EOM2}, the above equations define a non-hermitian system (non-unitary dynamics) of the field $\psi$ and the fields $\phi_n$.
The above system may be written as follows:
\begin{align} 
\I\partial_t\Psi(x,t)=M(x-x',t)\Psi(x',t),
\end{align}
where $\Psi(x,t)=\Big(\psi(x,t) \,\,...\,\,\phi_{-1}(x,t)\,\,\,\phi_{0}(x,t)\,\,\,\phi_{1}(x,t)\,\,...\Big)^T$, and $M$ is the non-hermitian Hamiltonian of the system. 

A knowledge of the field $\psi$ and the fields $\phi_n$ at $t=0$, determines completely the dynamics, i.e.,
\begin{align} \label{dynm}
\Psi(x,t)=\mathcal{T}\exp{\Big(-\I\int_0^t M(x-x',t')\mathrm{d}t'\Big)}\Psi(x',0).
\end{align}

Considering a wave-packet with $x$ and $n$ as coordinates. At any time, the center of the wave-packet, both in $x$ and $n$ spaces, may be evaluated:
\begin{align} 
&\langle x(t)\rangle=\int\mathrm{d}x'\,x|\psi(x,t)|^2,\\ \nonumber
& \langle n(t)\rangle=\sum n |\phi_n(x,t)|^2,
\end{align}

For the Hamiltonian and the memory kernel of the $1+1$D case in the main text, we now consider four different initial states. The first two states are localized in the middle of the chain with a well defined number of photons, where the topological transitions occur:
\begin{align} 
\Psi_{1,\pm}(x,0)=\begin{cases}
\psi(x,0)=\phi_{n=\pm n_1}(x,0)=e^{-(x/x_0)^2}\left(\begin{matrix}
    1 \\
    0
		\end{matrix}\right)\\
\phi_{n\neq \pm n_1}(x,0)=0.
\end{cases}
\end{align}
The other two states are localized on the right/left edge of the chain with no photons:
\begin{align} 
\Psi_{2,\pm}(x,0)=\begin{cases}
\psi(x,0)=\phi_{n=0}(x,0)=\delta(x\pm L/2)\left(\begin{matrix}
    1 \\
    0
		\end{matrix}\right)\\
\phi_{n\neq 0}(x,0)=0.
\end{cases}
\end{align}

\begin{figure}[t]
\centering
\includegraphics[width =\linewidth]{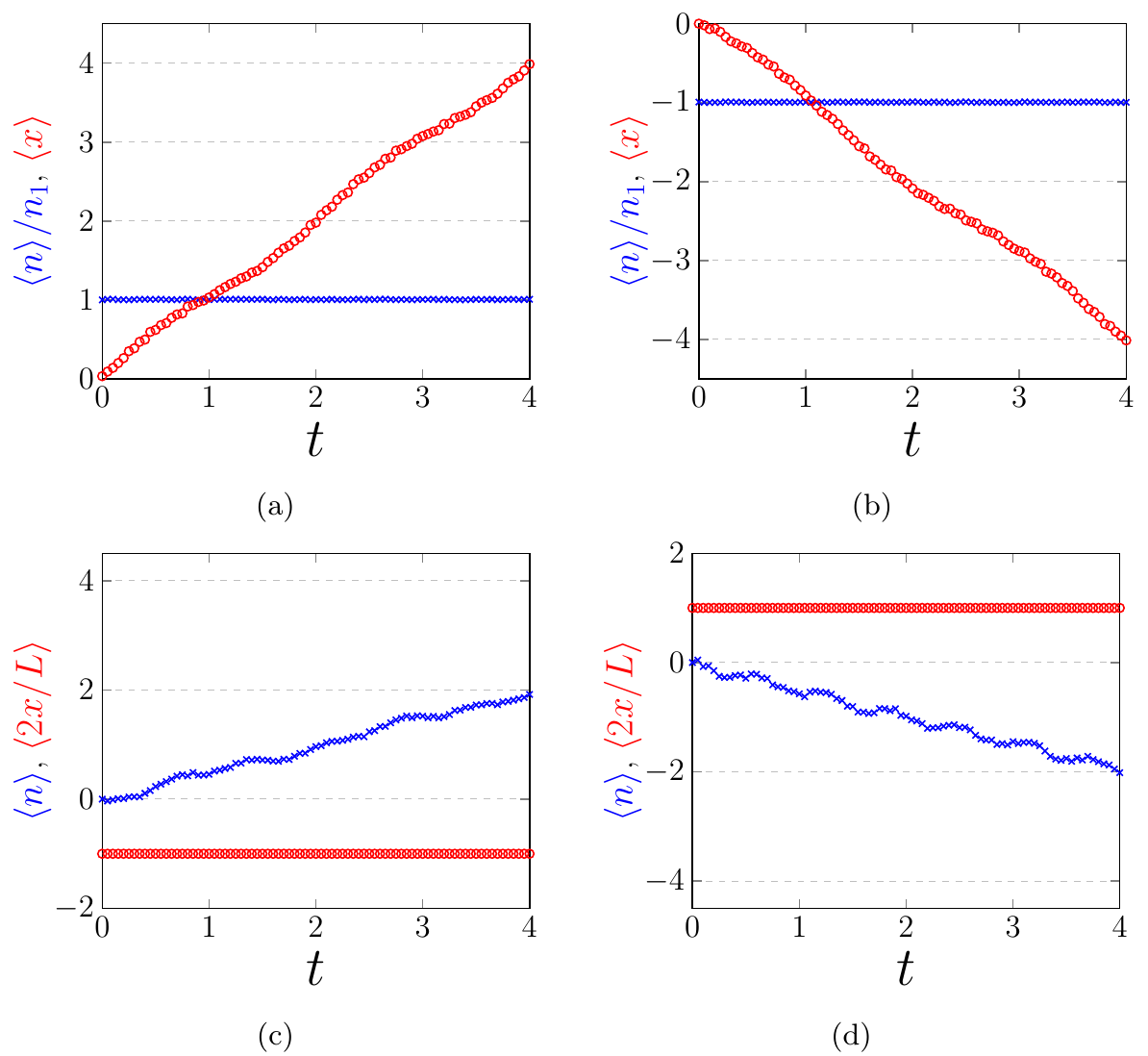} 
\caption{The center of a wave-packet in real space, $\langle x(t)\rangle$, (red) and in photon space, $\langle n(t)\rangle$, (blue) as a function of time for the four different initial states: (a) $\psi_{1,+}$ (b) $\psi_{1,-}$ (c) $\psi_{2,+}$ (d) $\psi_{2,-}$. \label{num}}
\end{figure}

In all four cases we used \refEq{dynm} to determine the time evolution and calculate $\langle x(t)\rangle$ and $\langle n(t)\rangle$. In all cases we avoided the corners in the $x-n$ space where the numerics is not reliable because of the sharp edges. In the following, we set $v_0$ and $\omega$ to unity and therefore $T=2\pi$. We find the evolution of the initial states using \refEq{dynm} for short time segments $dt=10^{-4}T$. See \refFig{num}. Evidently, the numerical results agree qualitatively with the analytical analysis. 

\section{Appendix B: Pseudo-spin field coupled to a chiral field}
\begin{figure}[t]
\centering
\includegraphics[width =\linewidth]{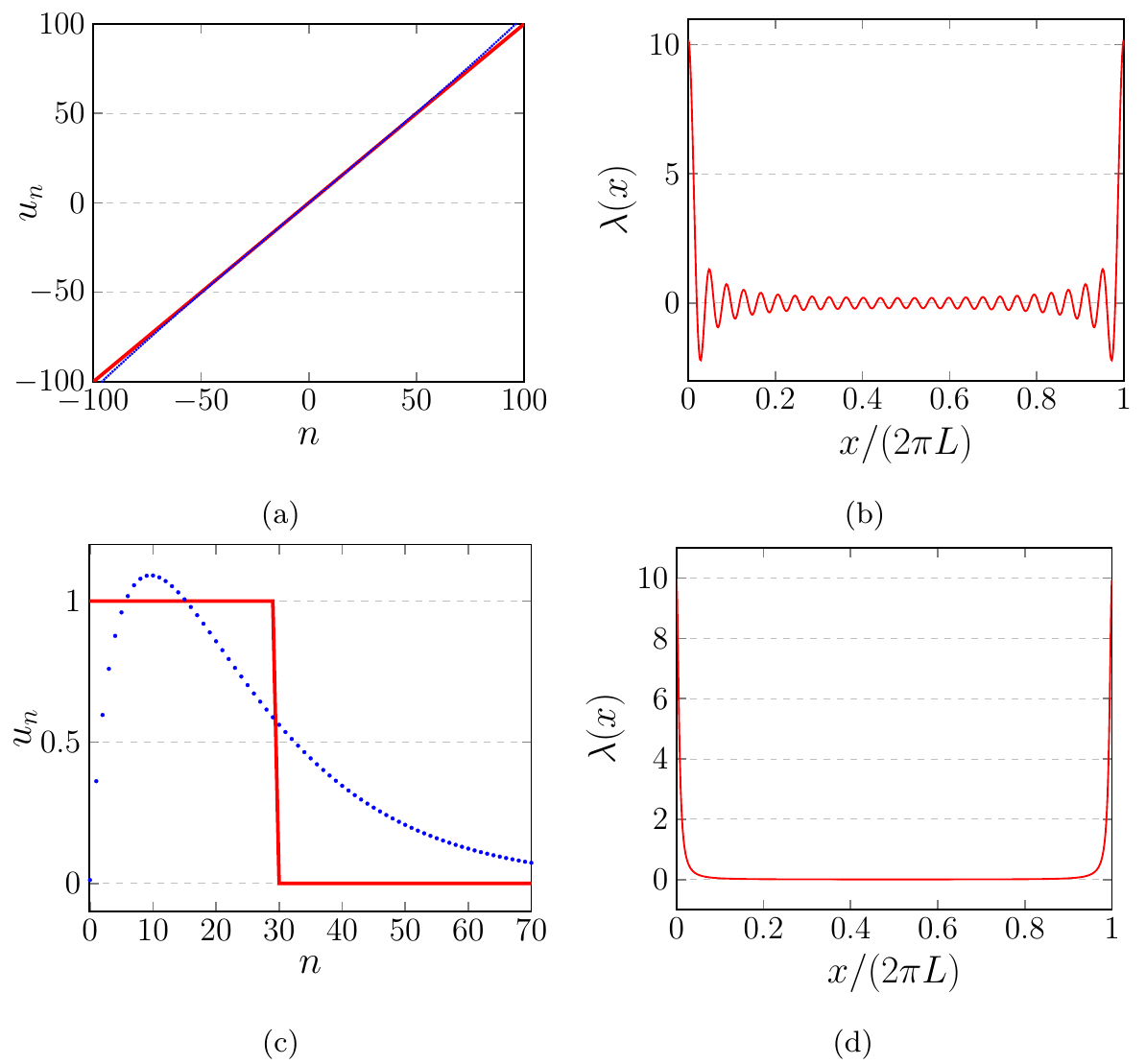} 
\caption{For $\lambda_n=\mbox{const}\times\Theta(|n|-250)$ (a) is the resultant $u_n$ (blue) vs. a perfect linear potential (red) and (b) is the corresponding $\lambda(x)$. (c,d) same as (a,b) only for $\lambda_n=\exp{(-|n|/25)}$.\label{coupling_spec}}
\end{figure}

In the main text we discussed the case of a (pseudo) spin field coupled to a chiral field and showed that a memory kernel of the desired form may be achieved. In \refEq{kernel_real} we found that the Fourier components of the memory kernel $u(t)$ may be expressed in terms of the Fourier components of the coupling $\lambda(x)$ as follows:
\begin{align}\nonumber
u_n=\sum\limits_{m\neq n}\frac{2\mbox{Re}(\lambda_{m}\,\lambda_{n}^{\dagger})}{2\pi (n-m)}.
\end{align}
if $\lambda_n=\mbox{const}$ for $|n|<n_0$ and zero otherwise, the resultant $u_n$ is approximately linear for $|n|\ll n_0$. For large $n_0$ this is sufficient to screen the electric field in an extensive region. These $\lambda_l$ correspond to the function shown in \refFig{coupling_spec}b, which is oscillatory with wave number $n_0/L$. The resultant $u_n$ and the corresponding $\lambda(x)$ are depicted in \refFig{coupling_spec}a and b.
An exact rectangular function is not accessible, but a smoother version of a rectangular function can be generated. For example, for $\lambda_n\sim\exp{(-|n|/n_1)}$, a smooth crossover between a topological and trivial region occurs. The resultant $u_n$ and the corresponding $\lambda(x)$ are depicted in \refFig{coupling_spec}c and d.

In the main text we consider the coupling of a spin field to a chiral $1$D field. However, it is possible to get a similar effect with a non-chiral field. 
Assume a standard non-chiral wire (quadratic dispersion) in which the chemical potential is located high above the bottom of the band. The excitations in that case can be treated as two decoupled (left and right) chiral fields.
In that case the corresponding memory kernel has the required form and its Fourier components are given by:
\begin{align}\label{kernel_real_non_chiral}
u_n=\sum\limits_{m\neq n}\frac{\mbox{Re}(\lambda_{R,m}\,\lambda_{R,n}^{\dagger}+\lambda_{L,-m}\,\lambda_{L,-n}^{\dagger})}{\pi (n-m)},
\end{align}
where $\lambda_{R/L}$ refer to the coupling to the right/left moving modes. This formula may lead to similar memory kernels as in the chiral case. For clean systems, deviations from the above formula arise due to curvature effects. However, as long as the chemical potential is the largest energy scale in the problem, these deviations are small.


\begin{thebibliography}{43}
\bibitem [{tra()}]{trans_com}%
  \BibitemOpen
  \href@noop {} {}\bibinfo {note} {In practice, also the linear potential
  corresponds to a uniform force and hence may be replaced by a uniform
  time-dependent vector potential.}\BibitemShut {Stop}%
\bibitem [{act()}]{action}%
  \BibitemOpen
  \href@noop {} {}\bibinfo {note} {The EOM can be derive from the following
  action: $
  \mathcal{S}=\int{\mathrm{d}t\phi^{\dagger}(t)[\I\partial_t-H_{\phi}(t)]\phi(t)}+\int{\mathrm{d}t\mathrm{d}x\lambda(x)\phi^{\dagger}(t)\psi(x,t)}+h.c.+\int{\mathrm{d}t\mathrm{d}x\psi^{\dagger}(x,t)[\I\partial_t-H_{\psi}(\hat{x})]\psi(x,t)}
  $}\BibitemShut {NoStop}%
\end{thebibliography}
\end{document}